\documentclass[11pt]{article}

\usepackage[final]{acl}
\usepackage{amssymb}
\usepackage{booktabs}   
\usepackage{multirow}   
\usepackage{times}
\usepackage{latexsym}
\usepackage[T1]{fontenc}
\usepackage[utf8]{inputenc}
\usepackage{microtype}
\usepackage[T1]{fontenc}
\usepackage{amsmath}
\usepackage{amssymb}
\usepackage[utf8]{inputenc}
\usepackage{relsize}    
\usepackage{bbm} 

\usepackage{microtype}

\usepackage{graphicx}
\usepackage{booktabs}
\usepackage{multirow}
\usepackage{graphicx}
\usepackage{amsmath}
\usepackage{amssymb}
\usepackage{mathtools}
\usepackage{booktabs}
\usepackage{array}
\usepackage{multirow}
\usepackage{enumitem}
\usepackage{float}
\usepackage{listings}
\usepackage{tcolorbox}

\usepackage[ruled,linesnumbered]{algorithm2e}
\usepackage{setspace}
\title{Instructions for *ACL Proceedings}

\author{First Author \\
  Affiliation / Address line 1 \\
  Affiliation / Address line 2 \\
  Affiliation / Address line 3 \\
  \texttt{email@domain} \\\And
  Second Author \\
  Affiliation / Address line 1 \\
  Affiliation / Address line 2 \\
  Affiliation / Address line 3 \\
  \texttt{email@domain} \\}

\title{Seeing and Reflecting: Multimodal Memory-Enhanced Agent Collaboration for Recommendation}

\author{
 \textbf{Hao Cong\textsuperscript{1,*}},
 \textbf{Huizu Lin\textsuperscript{2,*}},
 \textbf{Zihan Wang\textsuperscript{3,*}},
 \textbf{Chengkai Huang\textsuperscript{4,5,\#}},\\
 \textbf{Quan Z. Sheng\textsuperscript{5}},
 \textbf{Lina Yao\textsuperscript{4,6}}
\\
\\
 \textsuperscript{1}Tsinghua University,
 \textsuperscript{2}University of Science and Technology of China,
\\
 \textsuperscript{3}Peking University,
 \textsuperscript{4}The University of New South Wales,
\\
 \textsuperscript{5}Macquarie University,
 \textsuperscript{6}CSIRO's Data61
\\
\\
 \small{\textsuperscript{*}Equal contribution. 
 \textsuperscript{\#}Corresponding author.}
\\
 \small{\textbf{Correspondence:} 
 \href{mailto:chengkai.huang1@unsw.edu.au}{chengkai.huang1@unsw.edu.au}}
}

\begin{document}
\maketitle

\begin{abstract}


Large language model (LLM)-based agentic recommender systems show promise in modeling user preferences through natural-language reasoning, yet they remain limited by text-centric inputs and coarse-grained memory updates, making agents prone to missing visual evidence, semantic noise, and preference drift. To address these limitations, we propose \textbf{MMEACR}, a \textbf{M}ultimodal \textbf{M}emory-\textbf{E}nhanced \textbf{A}gent \textbf{C}ollaboration framework for recommendation. MMEACR introduces a dual-track memory architecture that separates interpretable agent reasoning from fine-grained multimodal matching. In the reasoning track, collaborative User and Item Memory Agents maintain persistent multimodal memories and update them through an attribute-guided reinforcement-and-reflection mechanism. In the matching track, a decoupled multimodal embedding memory is built from raw interaction narratives and item images to preserve detailed cross-modal signals beyond structured memory updates. The two tracks are integrated through weighted Reciprocal Rank Fusion to produce robust and interpretable rankings. Experiments on three real-world domains show that MMEACR achieves strong overall performance against competitive LLM-based and agent-based baselines, with notable gains in visually grounded recommendation scenarios.
\end{abstract}

\section{Introduction}

Large language models (LLMs) have recently enabled a new class of agentic recommender systems that model user preferences through natural-language reasoning and interaction~\cite{huang2025towards,huang2024foundation,huang2025survey}. 
Representative methods such as AgentCF~\cite{agentcf} formulate recommendation as a collaborative process among language agents, where agents compare candidate items, generate rationales, and update their internal memories. 
This paradigm improves interpretability and flexibility by externalizing preference modeling into language-mediated reasoning rather than relying solely on gradient-based representation learning.

\begin{figure}[t] 
    \centering
    \includegraphics[width=\columnwidth]{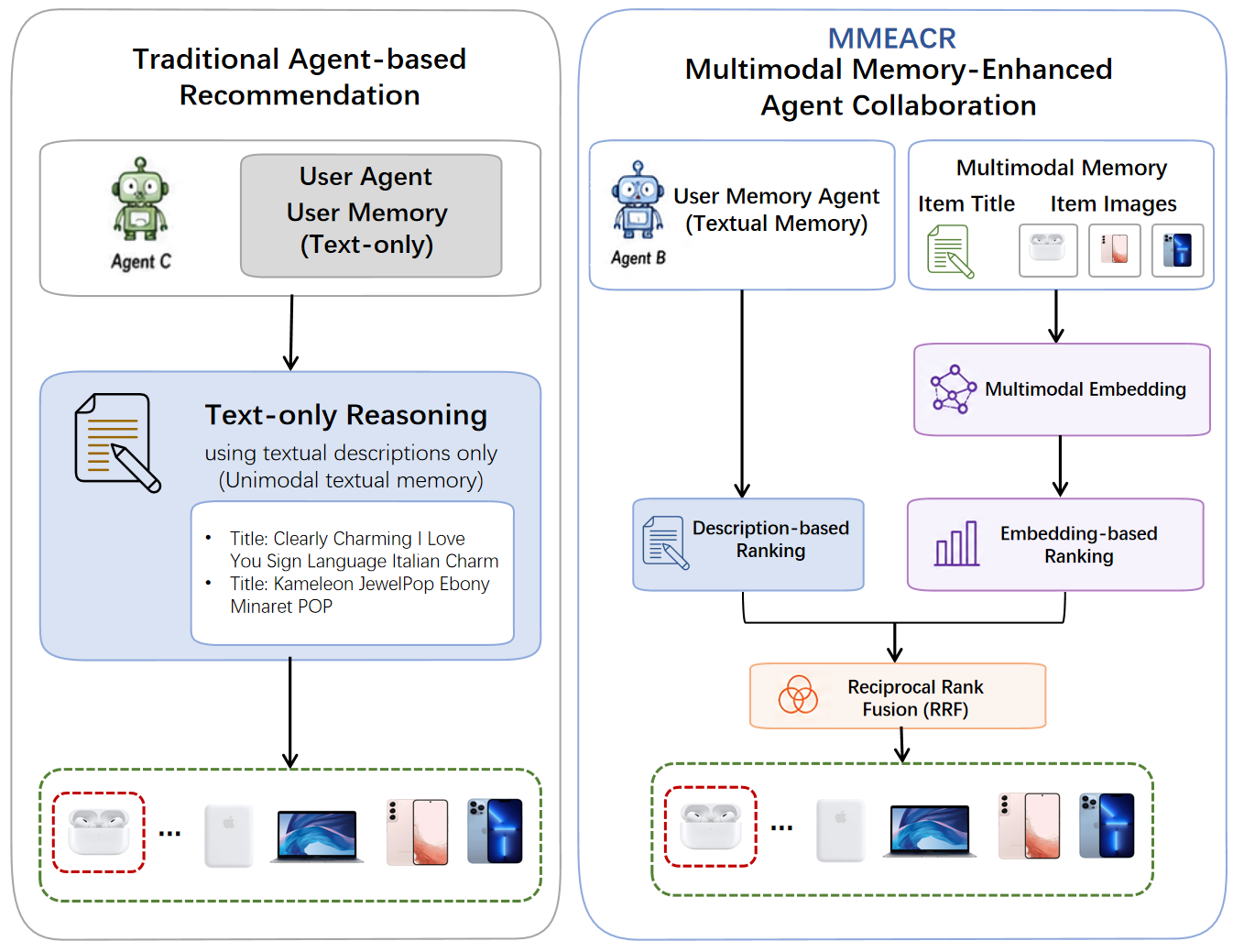} 
    \vspace{0.3em} 
    \parbox{0.43\columnwidth}{\centering \small (a) Traditional agent-based recommendation}
    \hfill
    \parbox{0.55\columnwidth}{\centering \small (b) MMEACR framework}
    \caption{Comparison between conventional text-centric agent-based recommendation and MMEACR. MMEACR combines multimodal agent memory evolution with embedding-based cross-modal matching and fuses their rankings via Reciprocal Rank Fusion (RRF).}
    \label{fig:framework_comparison}
\end{figure}

Despite their promise, existing LLM-based recommendation agents still suffer from two key limitations. 
First, most methods remain text-centric, constructing user and item memories mainly from textual histories, reviews, or metadata while underusing visual evidence such as product images~\cite{agentcf++}. 
This restricts their ability to model preferences in visually grounded domains such as fashion, electronics, and media products, where appearance, style, packaging, and other visual attributes often play an important role. 
Second, current memory update mechanisms are often coarse-grained. 
Agents typically revise their profiles through free-form reflection or direct history accumulation, which can introduce redundant descriptions, amplify spurious rationales, and cause preference drift over repeated interactions. 
These issues become more challenging when textual and visual signals must be jointly interpreted and selectively consolidated.

To address these challenges, we propose \textbf{MMEACR}, a \textbf{M}ultimodal \textbf{M}emory-\textbf{E}nhanced \textbf{A}gent \textbf{C}ollaboration framework for recommendation. 
MMEACR is motivated by the idea of \textit{Seeing and Reflecting}. 
\textit{Seeing} grounds agents in multimodal evidence by initializing and enriching item memories with both textual metadata and image-derived descriptions. 
\textit{Reflecting} enables structured memory evolution by reinforcing aligned preference signals and correcting misleading ones after user--item interactions.

MMEACR contains two complementary tracks. 
The reasoning track uses cooperative User and Item Memory Agents to maintain persistent language memories. 
During interaction, an LLM-based agent compares candidate items, generates a rationale, and updates memories according to whether the predicted preference matches the observed feedback. 
To reduce noisy updates, MMEACR extracts preference-relevant attributes from a predefined semantic attribute space and uses them to guide both reinforcement and reflection. 
This helps preserve stable user interests while revising inaccurate or drifting memory descriptions.

The matching track complements language-based reasoning with dense multimodal representations. 
Since structured memory updates may filter out fine-grained details, MMEACR also maintains raw interaction narratives and item images for embedding-based matching. 
A multimodal embedding module encodes these signals into dense representations, capturing cross-modal semantics that may be difficult to express in concise textual memories. 
Finally, MMEACR combines the reasoning-based and embedding-based rankings through weighted Reciprocal Rank Fusion, allowing the final recommendation to benefit from both interpretable agent reasoning and fine-grained multimodal similarity.


Our contributions are summarized as follows:
\begin{itemize}
    \item We propose \textbf{MMEACR}, a multimodal memory-enhanced agent collaboration framework that grounds LLM-based recommendation agents in both textual and visual evidence.
    \item We design an attribute-guided memory evolution mechanism that enables User and Item Memory Agents to reinforce aligned preferences and reflect on misaligned ones, reducing semantic noise and preference drift.
    \item We introduce a dual-track recommendation pipeline that combines interpretable language-agent reasoning with multimodal embedding-based matching through Reciprocal Rank Fusion, achieving strong performance across multiple domains.
\end{itemize}

\section{Related Work}

\subsection{LLM-based Recommendation}
Large language models (LLMs) have recently been explored for recommendation due to their strong abilities in semantic understanding, instruction following, and natural-language reasoning~\cite{hou2024llm,zhao2024recommender,gao2026factorized}. 
Early LLM-based recommenders mainly use frozen LLMs as zero-shot or few-shot rankers, where user histories and candidate items are converted into textual prompts for preference prediction~\cite{hou2024llm}. 
Other studies further improve task adaptation through Chain-of-Thought reasoning, instruction tuning, or retrieval-augmented generation, enabling LLMs to better align textual semantics with collaborative signals~\cite{wei2022chain,bao2023tallrec,shi2025retrieval}. 
More recently, agentic recommender systems have represented users and items as autonomous language agents that interact, exchange feedback, and update their memories during recommendation~\cite{agentcf,agentcf++}. 
These methods improve interpretability by making preference modeling explicit in language-based reasoning processes. 
Meanwhile, multimodal LLMs and multimodal embedding models have shown the potential to incorporate visual and textual signals for richer item understanding~\cite{macawllm2023macaw,gme}. 
However, most existing agentic recommenders still rely primarily on textual memories or directly inject multimodal information into prompts, which may lead to verbose contexts, redundant semantics, and unstable reasoning. 
In contrast, MMEACR adopts a dual-track design that separates structured agentic reasoning from dense multimodal matching, allowing the framework to use visual evidence without overloading the language-agent memory.

\subsection{Memory in LLM Agents}
Memory is essential for LLM agents to maintain consistency, accumulate experience, and support reflection across interactions~\cite{ferrag2025from,zhang2025survey,ye2026memweaver}. 
Existing agent memory mechanisms range from short-term context buffers to long-term episodic stores, retrieval-based memories, and reflection-based memory management~\cite{tang2026llm,jiao2026doctor,jiao2026prunerag}. 
Although these mechanisms have been widely studied in general-purpose agent systems, applying them to recommendations remains challenging. 
User preferences are often dynamic, sparse, and attribute-dependent, while item descriptions may contain noisy or redundant information. 
Simply appending interaction histories or retrieving past episodes can therefore introduce irrelevant details, amplify spurious preference signals, and cause preference drift over time~\cite{salve2024collaborative,li2025clear}. 
Recent memory-enhanced recommender agents attempt to update user and item profiles through interaction feedback, but their updates are often coarse-grained and lack explicit constraints on which semantic attributes should be reinforced or corrected. 
MMEACR addresses this limitation with an attribute-guided memory evolution mechanism, where preference-relevant attributes are extracted from user-item comparisons and used to guide reinforcement and reflection. 
This design enables agents to consolidate stable preference signals while reducing semantic noise in long-term multimodal memory.

\section{Problem Formulation}

Let $\mathcal{U}$ and $\mathcal{I}$ denote the sets of users and items, respectively. 
For each user $u \in \mathcal{U}$, we assume a chronologically ordered interaction history that reflects the user's past preferences. 
Given a candidate item set $\mathcal{C} = \{c_1, \dots, c_n\} \subseteq \mathcal{I}$, the goal is to generate a personalized ranking of these candidates according to the user's preference. 
Formally, we define the recommendation function as:
\begin{equation}
    \hat{\mathcal{C}} = f_{\mathrm{LLM}}(u, \mathcal{C}),
\end{equation}
where $\hat{\mathcal{C}}$ denotes the ranked candidate list produced by the LLM-based recommendation framework.

\noindent \textbf{Memory in Agents.}
Following the agent-based recommendation paradigm~\cite{agentcf}, we model the collaborative recommendation process with two types of LLM-powered memory agents: a \textbf{User Memory Agent (UA)} and an \textbf{Item Memory Agent (IA)}. 
Rather than relying on static textual profiles or fixed representation embeddings, MMEACR treats memories as persistent semantic states that can be updated through interactions. 
Specifically, the user memory $M_u$ is an evolving textual narrative that summarizes the user's long-term preferences, behavioral patterns, and fine-grained attribute tendencies. 
The item memory $M_i$ describes item $i$ by integrating its textual metadata with visual descriptions derived from product images. 
During recommendation and memory evolution, the User and Item Memory Agents collaborate through LLM-based reasoning over $\{M_u\} \cup \{M_i\}$, enabling the framework to capture preference-relevant semantic signals and support reasoning-aware user--item interactions.

\section{Methodology}

\begin{figure*}[t]
    \centering
    \includegraphics[width=\linewidth]{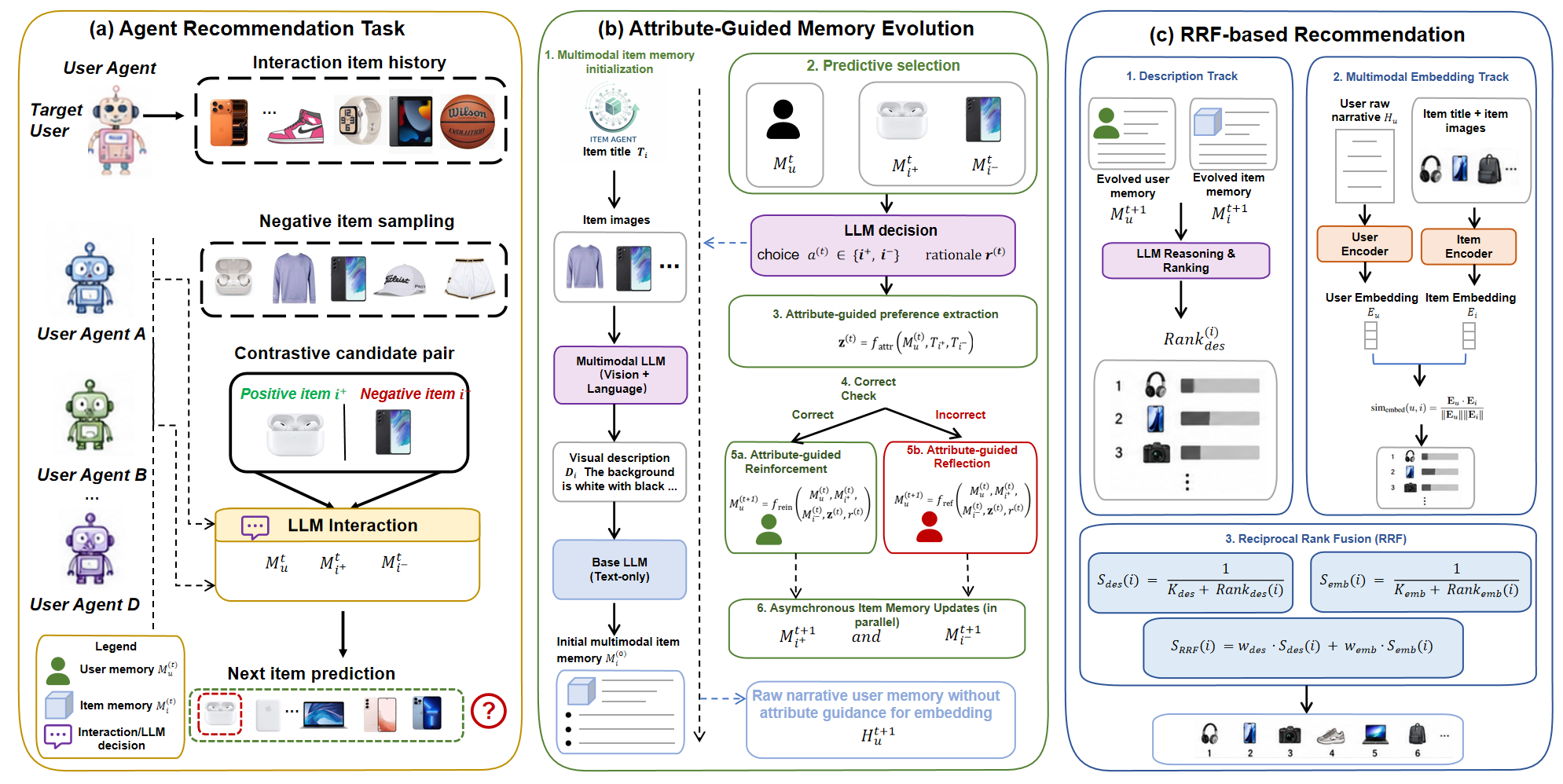}
    \vspace{-1em}
    \caption{Overview of MMEACR. The reasoning track performs attribute-guided memory evolution with User and Item Memory Agents, while the matching track preserves fine-grained cross-modal signals through multimodal embedding memory. The final ranking is produced by fusing both tracks via RRF.}
    \label{fig:overview}
    \vspace{-0.5em}
\end{figure*}

We present \textbf{MMEACR}, a multimodal memory-enhanced agent collaboration framework for recommendation.
As shown in Figure~\ref{fig:overview}, MMEACR contains two complementary tracks: 
(1) a \textit{reasoning track}, where User and Item Memory Agents maintain and update interpretable multimodal memories through LLM-based interaction; and 
(2) a \textit{matching track}, where raw interaction narratives and item images are encoded into dense multimodal embeddings.
The two tracks are finally combined through weighted Reciprocal Rank Fusion (RRF) to produce the final recommendation ranking.

\subsection{Multimodal Agent Memory}

\subsubsection{Memory Initialization}

MMEACR first initializes semantic memories for users and items.
For each item $i \in \mathcal{I}$, let $T_i$ denote its textual title and $\mathcal{V}_i=\{v_{i,1},\dots,v_{i,N_i}\}$ denote its associated images, where $N_i \leq 5$.
To convert visual evidence into language-readable semantics, we use a multimodal LLM to generate an image-grounded description:
\begin{equation}
    D_i = \mathrm{MLLM}(T_i, \mathcal{V}_i).
\end{equation}
The initial item memory is then constructed by integrating the title and visual description:
\begin{equation}
    M_i^{(0)} = \mathrm{LLM}(T_i, D_i).
\end{equation}
Here, $M_i^{(0)}$ serves as a multimodal textual memory that summarizes the item's key attributes.
For each user $u$, we initialize the user memory $M_u^{(0)}$ with a domain-specific template, such as a general preference statement for music, electronics, or fashion products.
In addition to structured memories, we maintain raw narratives $H_u$ and $H_i$ for users and items, which preserve unfiltered interaction histories and multimodal descriptions for embedding-based matching.

\subsubsection{Attribute-Guided Memory Evolution}

The reasoning track updates agent memories through contrastive user-item interactions.
At each interaction step $t$, we construct a triplet $(u,i^+,i^-)$, where $i^+$ is the ground-truth preferred item and $i^-$ is a sampled negative item.
Given the current user memory $M_u^{(t)}$ and item memories $M_{i^+}^{(t)}$ and $M_{i^-}^{(t)}$, the LLM-based interaction agent selects the item that better matches the user's preference and generates a rationale:
\begin{equation}
    a^{(t)}, r^{(t)} =
    f_{\mathrm{dec}}\left(M_u^{(t)}, M_{i^+}^{(t)}, M_{i^-}^{(t)}\right),
\end{equation}
where $a^{(t)} \in \{i^+, i^-\}$ is the selected item and $r^{(t)}$ is the natural-language rationale.
The correctness of the selection is determined by:
\begin{equation}
    y^{(t)} = \mathbb{I}\left[a^{(t)} = i^+\right].
\end{equation}

To avoid updating memories with noisy free-form rationales, MMEACR introduces an attribute-guided preference extraction step.
Let $\mathcal{A}=\{a_1,\dots,a_K\}$ be a predefined semantic attribute space, where each attribute corresponds to a high-level preference factor such as style, functionality, portability, or visual appearance.
Conditioned on the current memory, the contrastive item pair, and the generated rationale, the attribute extractor produces a structured preference signal:
\begin{equation}
    \mathbf{z}^{(t)} =
    f_{\mathrm{attr}}\left(M_u^{(t)}, T_{i^+}, T_{i^-}, r^{(t)}, \mathcal{A}\right),
\end{equation}
where $\mathbf{z}^{(t)} \in \mathbb{R}^{K}$ encodes the attributes that explain the preference difference between $i^+$ and $i^-$.

The extracted attribute signal is then used to guide memory evolution.
If the LLM selection is correct, MMEACR reinforces the current preference pattern; otherwise, it performs reflection to correct the misaligned memory:
\begin{equation}
\small
\begin{aligned}
\mathbf{x}_u^{(t)} &= 
\left(M_u^{(t)}, M_{i^+}^{(t)}, M_{i^-}^{(t)}, \mathbf{z}^{(t)}, r^{(t)}\right), \\
M_u^{(t+1)} &=
\begin{cases}
f_{\mathrm{rein}}\!\left(\mathbf{x}_u^{(t)}\right), & y^{(t)}=1, \\
f_{\mathrm{refl}}\!\left(\mathbf{x}_u^{(t)}\right), & y^{(t)}=0.
\end{cases}
\end{aligned}
\end{equation}
This design allows the User Memory Agent to consolidate stable interests while revising inaccurate or drifting preference descriptions.

The Item Memory Agents are updated asynchronously for the interacted positive and negative items:
\begin{align}
    M_{i^+}^{(t+1)} &= 
    f_{\mathrm{pos}}\left(M_{i^+}^{(t)}, M_u^{(t)}, \mathbf{z}^{(t)}, y^{(t)}\right), \\
    M_{i^-}^{(t+1)} &= 
    f_{\mathrm{neg}}\left(M_{i^-}^{(t)}, M_u^{(t)}, \mathbf{z}^{(t)}, y^{(t)}\right).
\end{align}
Meanwhile, the raw user narrative is updated by appending the preferred item's metadata and visual description:
\begin{equation}
    H_u^{(t+1)} = H_u^{(t)} \oplus [T_{i^+}; D_{i^+}],
\end{equation}
where $\oplus$ denotes chronological concatenation.
Similarly, raw item narratives are maintained for the interacted items.
These raw narratives are not filtered by the attribute space, ensuring that fine-grained cross-modal information remains available for the embedding-based matching track.

\subsubsection{Reasoning-Based Ranking}

After iterative memory evolution, the reasoning track ranks candidate items using the evolved user and item memories.
Given a candidate set $\mathcal{C}$, the LLM-based ranker compares the user memory $M_u$ with the candidate item memories $\{M_c \mid c \in \mathcal{C}\}$ and produces a description-based ranking:
\begin{equation}
    \pi_{\mathrm{des}} =
    \mathrm{Rank}_{\mathrm{LLM}}\left(M_u, \{M_c \mid c \in \mathcal{C}\}\right).
\end{equation}
This ranking is interpretable because it is derived from explicit user and item memory descriptions.

\subsection{Multimodal Embedding Memory}

While the reasoning track provides interpretable preference reasoning, structured memory updates may filter out fine-grained visual or textual details.
Therefore, MMEACR introduces a multimodal embedding memory to preserve raw cross-modal signals for dense matching.

\subsubsection{Item Embedding Memory}

For each item $i$, we use a pretrained multimodal embedding model $\mathrm{MEM}$~\cite{gme} to encode its title and images.
The item embedding is computed by averaging image-title representations:
\begin{equation}
    \mathbf{e}_i =
    \frac{1}{N_i}\sum_{j=1}^{N_i}
    \mathrm{MEM}(T_i, v_{i,j}).
\end{equation}
This embedding captures both textual and visual item semantics.

\subsubsection{User Embedding Memory}

For each user $u$, we construct the user embedding from the raw interaction narrative and the images of historically preferred items.
Let $\mathcal{P}_u$ denote the set of items preferred by user $u$ in the interaction history.
The user embedding is computed as:
\begin{equation}
    \mathbf{e}_u =
    \frac{1}{|\mathcal{P}_u|}
    \sum_{i \in \mathcal{P}_u}
    \mathrm{MEM}(H_u, \bar{v}_i),
\end{equation}
where $\bar{v}_i$ denotes the representative visual content of item $i$.
Unlike the attribute-guided memory $M_u$, the raw narrative $H_u$ preserves complete historical descriptions, allowing the embedding track to capture detailed semantic and visual preference signals.

\subsubsection{Embedding-Based Ranking}

The embedding track ranks candidate items by cosine similarity:
\begin{equation}
    s_{\mathrm{emb}}(u,i) =
    \frac{\mathbf{e}_u^\top \mathbf{e}_i}
    {\|\mathbf{e}_u\| \|\mathbf{e}_i\|}.
\end{equation}
Sorting candidates by $s_{\mathrm{emb}}(u,i)$ yields the embedding-based ranking $\pi_{\mathrm{emb}}$.

\subsection{Hybrid Ranking via Reciprocal Rank Fusion}

The reasoning and matching tracks capture complementary signals.
The reasoning track provides interpretable preference judgments based on evolved agent memories, while the embedding track captures fine-grained cross-modal similarity from raw narratives and images.
To combine them, MMEACR adopts a weighted Reciprocal Rank Fusion strategy.

For each candidate item $i$, we first compute its rank-based scores from the two tracks:
\begin{align}
    S_{\mathrm{des}}(i) &=
    \frac{1}{k_{\mathrm{des}} + \mathrm{rank}_{\pi_{\mathrm{des}}}(i)}, \\
    S_{\mathrm{emb}}(i) &=
    \frac{1}{k_{\mathrm{emb}} + \mathrm{rank}_{\pi_{\mathrm{emb}}}(i)},
\end{align}
where $\mathrm{rank}_{\pi_{\mathrm{des}}}(i)$ and $\mathrm{rank}_{\pi_{\mathrm{emb}}}(i)$ denote the positions of item $i$ in the reasoning-based and embedding-based rankings, respectively.
The constants $k_{\mathrm{des}}$ and $k_{\mathrm{emb}}$ control the smoothness of rank contributions.

The final fusion score is:
\begin{equation}
    S_{\mathrm{RRF}}(i) =
    w_{\mathrm{des}} S_{\mathrm{des}}(i)
    + w_{\mathrm{emb}} S_{\mathrm{emb}}(i),
\end{equation}
where $w_{\mathrm{des}}$ and $w_{\mathrm{emb}}$ balance the contributions of the two tracks.
The final recommendation list is obtained by sorting all candidate items in descending order of $S_{\mathrm{RRF}}(i)$.

\begin{table*}[htbp]
\small
\centering
\caption{Performance comparison on three domains. Best results are in \textbf{bold}, second-best are \underline{underlined}. \textit{Imp.} denotes the relative improvement of our method over the strongest baseline. All results are averaged over five runs with different seeds.}
\vspace{-1em}
\label{tab:main_result}
\resizebox{\textwidth}{!}{
\begin{tabular}{lcccc|cccc|cccc}
\toprule
\multirow{2}{*}{Method} & 
\multicolumn{4}{c}{CDs\_and\_Vinyl} & 
\multicolumn{4}{c}{Cell\_Phones\_and\_Accessories} & 
\multicolumn{4}{c}{Fashion} \\
\cmidrule(lr){2-5} \cmidrule(lr){6-9} \cmidrule(lr){10-13}
 & N@1 & N@5 & N@10 & MRR & N@1 & N@5 & N@10 & MRR & N@1 & N@5 & N@10 & MRR \\
\midrule
Pop & 0.1100 & 0.3237 & 0.4708 & 0.3138 & 0.1300 & 0.3173 & 0.4734 & 0.3177 & 0.1100 & 0.3241 & 0.4749 & 0.3180 \\
BM25 & 0.1600 & 0.1769 & 0.4439 & 0.2852 & 0.2100 & 0.3691 & 0.4747 & 0.3741 & 0.2100 & 0.3708 & 0.4351 & 0.3650 \\
SASRec & 0.1400 & 0.3206 & 0.4790 & 0.3253 & 0.1200 & 0.3096 & 0.4610 & 0.3020 & 0.1300 & 0.2934 & 0.4609 & 0.3034 \\
\midrule
LLMSeqSim & 0.1800 & 0.4157 & 0.5319 & 0.3897 & 0.2100 & 0.4267 & 0.5344 & 0.3938 & 0.1800 & \underline{0.4364} & \underline{0.5390} & 0.3982 \\
MLLMSeqSim & 0.1800 & 0.4541 & 0.5418 & 0.4005 & 0.1300 & 0.3226 & 0.4778 & 0.3234 & 0.1600 & 0.3844 & 0.5067 & 0.3573 \\
LLMRank   & 0.1895 & 0.3628 &0.5085 &0.3624 &0.2043 &0.4135 &0.5299 &0.3888& \underline{0.2200} &0.4014 &0.5359 &\underline{0.3987}  \\
AgentCF &  0.1900 & 0.3941 & 0.5169 & 0.3731 & 0.2000 & \underline{0.4393} & \underline{0.5496} & 0.4121 & 0.1700 & 0.3399 & 0.4943 & 0.3447 \\
\midrule
CoTAgent& \underline{0.2700} & \textbf{0.5941} & \underline{0.6315} & \underline{0.5127} & \underline{0.2800} & 0.4334 & 0.5339 & \underline{0.4213} & 0.1900 & 0.3783 & 0.5081 & 0.3772 \\
\midrule
MMEACR-DES & 0.3200 & 0.5157 & 0.6088 & 0.4900 & 0.3000 & 0.5066 & 0.5926 & 0.4692 &  0.2500 & 0.4641 & 0.5708 & 0.4407 \\
MMEACR-EMB & 0.2600 & 0.5298 & 0.6074 & 0.4847 & 0.2500 & 0.4979 & 0.5815 & 0.4517 & \textbf{0.3400} &  0.5276 &  0.6179 &  0.5019 \\
MMEACR-RRF & \textbf{0.3400} & \underline{0.5632} & \textbf{0.6354} & \textbf{0.5224} & \textbf{0.3200} & \textbf{0.5327} & \textbf{0.6214} & \textbf{0.5049} & 0.3200 & \textbf{0.5382} & \textbf{0.6230} & \textbf{0.5069} \\
\midrule
\textbf{Improv(\%)} & \textbf{20.59\%} &  \textbf{-5.20\%} &  \textbf{0.62\%} &  \textbf{1.89\%} &  \textbf{14.3\%} &  \textbf{21.26\%} &  \textbf{13.06\%} &  \textbf{19.84\%} &  \textbf{45.45\%} &  \textbf{23.33\%} &  \textbf{15.58\%} &  \textbf{27.14\%} \\
\bottomrule
\end{tabular}
}
\end{table*}

\section{Experiments}
\begin{table}[t]
\centering
\caption{Dataset statistics.}
\vspace{-1em}
\small
\label{tab:dataset_stats}
\begin{tabular}{lrrrr}
\toprule
\textbf{Data} & \textbf{\#Users} & \textbf{\#Items} & \textbf{\#Inter.} & \textbf{Sparsity} \\
\midrule
CDs & 100 & 781 & 500 & 99.36\% \\
Cell\_Phones & 100 & 753 & 500 & 99.34\% \\
Fashion & 100 & 763 & 500 & 99.34\% \\
\bottomrule
\end{tabular}
\vspace{-0.8em}
\end{table}

\textbf{Datasets.} Following previous work \cite{agentcf,huang2023modeling,huang2023dual,huang2024dual}, we conduct experiments on three text-intensive subsets of the Amazon review dataset (CDs, Cell\_phones, and Fashion).



\textbf{Baselines.} We evaluate our method against several representative baselines, including popularity-based \textbf{Pop}, text-matching \textbf{BM25} \cite{bm25},  self-attention sequential recommender \textbf{SASRec} ~\cite{kang2018selfEMNLP}
, LLM-based ranker \textbf{LLMRank} \cite{hou2024large}, embedding-similarity models \textbf{LLMSeqSim} \cite{harte2023leveraging} and \textbf{MLLMSeqSim}, as well as the agent-based \textbf{AgentCF} \cite{agentcf} and \textbf{CoTAgent}\cite{wei2022chain}. Specifically, \textbf{CoTAgent} incorporates zero-shot Chain-of-Thought (CoT) prompting to facilitate transparent decision-making and produce interpretable reasoning rationales for each recommendation. Pop ranks items by interaction frequency, while BM25 retrieves candidates using textual similarity to user histories. 
LLMRank uses GPT-4o-mini as a zero-shot ranker conditioned on sequential histories. LLMSeqSim constructs a session representation via LLM-generated item embeddings and retrieves top-k similar items, whereas MLLMSeqSim extends this with multimodal LLM embeddings. AgentCF treats users and items as autonomous LLM agents that interact and update memories to perform collaborative filtering.
For our method \textbf{MMEACR}, we additionally include three variants. 
\textbf{MMEACR-DES} uses the user's and item's latest memories with a 1-positive-9-negative candidate set, prompting the LLM to return a ranking for computing NDCG and MMR. 
\textbf{MMEACR-EMB} fuses the user's five recent item-image embeddings with the latest memory via GME, combines item image--title embeddings, and ranks candidates using cosine similarity. 
\textbf{MMEACR-RRF} merges the rankings from MMEACR-DES and MMEACR-EMB using the RRF formula.

\begin{table}[t]
\centering
\caption{Ablation study on Cell\_phones and Fashion. Best results are in \textbf{bold}, and second-best results are \underline{underlined}.}
\label{tab:ablation_study}
\small
\setlength{\tabcolsep}{5pt}
\renewcommand{\arraystretch}{1.08}
\begin{tabular*}{\columnwidth}{@{\extracolsep{\fill}}lcccc@{}}
\toprule
\multicolumn{5}{c}{\textbf{Cell\_phones}} \\
\midrule
Variant & N@1 & N@5 & N@10 & MRR \\
\midrule
Full 
& \textbf{.3000} & \textbf{.5066} & \underline{.5926} & \underline{.4692} \\
w/o Attr. 
& \underline{.2700} & \textbf{.5066} & \textbf{.5987} & \textbf{.4749} \\
w/o Refl. \& Attr. 
& .2200 & .4743 & .5663 & .4323 \\
w/o User \& Attr. 
& .1800 & .4047 & .5235 & .3804 \\
w/o Item \& Attr. 
& .2500 & \underline{.5051} & .5800 & .4500 \\
\midrule
\multicolumn{5}{c}{\textbf{Fashion}} \\
\midrule
Variant & N@1 & N@5 & N@10 & MRR \\
\midrule
Full 
& \textbf{.2500} & \textbf{.4641} & \textbf{.5708} & \textbf{.4407} \\
w/o Attr. 
& \underline{.2300} & .4492 & \underline{.5605} & \underline{.4277} \\
w/o Refl. \& Attr. 
& \underline{.2300} & \underline{.4555} & .5585 & .4245 \\
w/o User \& Attr. 
& .1500 & .3245 & .4826 & .3305 \\
w/o Item \& Attr. 
& \underline{.2300} & .4317 & .5468 & .4107 \\
\bottomrule
\end{tabular*}
\vspace{-0.5em}
\end{table}

\textbf{Evaluation Metrics.} We adopt a leave-one-out strategy ~\cite{kang2018selfEMNLP}: for each user, the last interaction is used for testing and the second-to-last for validation. Following previous works~\cite{agentcf}, each evaluation case includes one positive item and 9 randomly sampled negatives from the same domain.

\begin{figure*}[t]
    \centering
    \includegraphics[width=\linewidth]{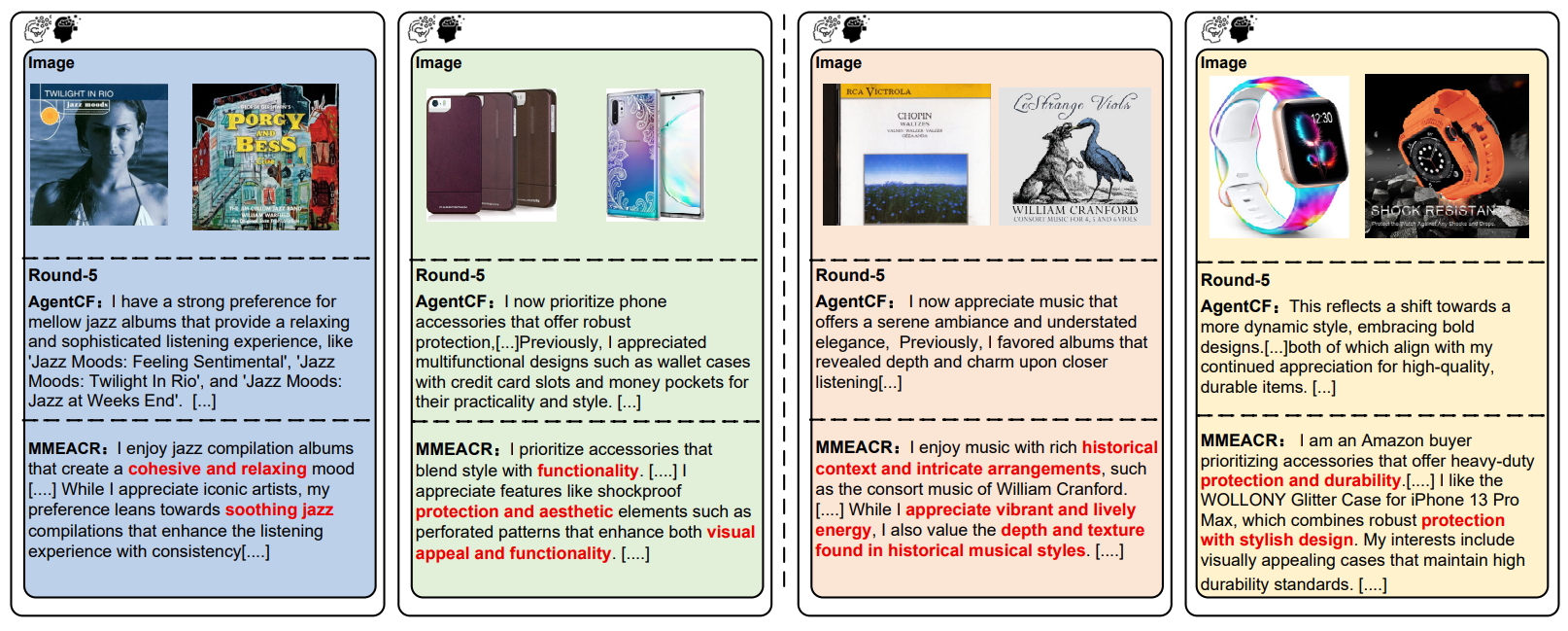}
    \caption{Case study of memory evolution and profile reflection at Round-5 across four semantic domains. Compared to AgentCF, our MMEACR demonstrates a superior capability to filter cross-modal noise and capture attribute-guided user preferences (highlighted in red) such as ``cohesive and relaxing mood'', ``protection and aesthetic'', and ``historical context'', securing more precise long-term interest alignment.}
    \label{fig:case}
\end{figure*}

\textbf{Quantative Analysis.} Table~\ref{tab:main_result} reports the overall comparison across three domains. 
Our proposed \textbf{MMEACR-RRF} consistently delivers the strongest or highly competitive performance across datasets and metrics. On \textit{CDs}, our method improves N@1 by 20.59\% and MRR by 1.89\% over the strongest baseline, while maintaining competitive performance on other metrics. The slightly lower NDCG@5 compared with CoTAgent may be attributed to the relatively low complexity of the CDs dataset, where explicit CoT reasoning is already sufficient to capture user preferences and rank relevant items effectively, leaving limited room for additional gains from attribute-guided reasoning and multimodal embedding fusion.
On \textit{Cell\_Phones}, the improvements are consistent across all metrics, with gains of 14.3\% on N@1 and 21.26\% on N@5, demonstrating robust ranking quality in more attribute-diverse scenarios. 
Notably, on the visually intensive \textit{Fashion} domain, our approach achieves substantial gains, improving N@1, N@5, and MRR by 45.45\%, 23.33\%, and 27.14\%, respectively, highlighting the benefit of multimodal memory modeling in visually grounded recommendation.

\begin{figure}[t] 
    \centering
\includegraphics[width=\columnwidth]{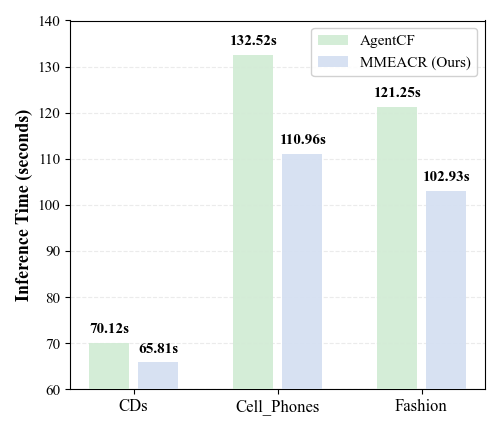}    
\caption{Inference time comparison between AgentCF and MMEACR on the three datasets. The Y-axis denotes the inference time (in seconds), while the X-axis denotes different datasets. MMEACR achieves faster inference across both datasets.}
    \label{compare_inference_time}
\end{figure}

Moreover, while both single-branch variants (MMEACR-DES and MMEACR-EMB) already achieve competitive performance, the fused model consistently yields further improvements, indicating that LLM-based reasoning and multimodal embedding similarity provide complementary signals. 
Compared with prior agent-based methods such as AgentCF and CoTAgent, our framework demonstrates consistent advantages across domains, validating the effectiveness of structured multimodal memory and reflective agent collaboration. To evaluate the computational efficiency of our proposed framework, we compare the inference time of MMEACR against the representative agent-based baseline, AgentCF, across three datasets. As illustrated in Figure \ref{compare_inference_time}, MMEACR consistently achieves lower inference latency than AgentCF in all evaluated domains. Specifically, MMEACR reduces inference time by 6.15\% (from 70.12s to 65.81s) on CDs, 16.27\% (from 132.52s to 110.96s) on Cell\_Phones, and 15.11\% (from 121.25s to 102.93s) on the Fashion dataset. These results demonstrate that while MMEACR incorporates structured multimodal memory and reflective collaboration to achieve superior ranking performance (as shown in Table \ref{tab:main_result}), it simultaneously optimizes the inference pipeline, proving its high efficiency and practicality for real-world recommendation scenarios.

\textbf{Qualitative Analysis.} As shown in Figure~\ref{fig:case}, while AgentCF generates coarse-grained and repetitive interest summaries (e.g., general ``jazz'' or ``phone accessories''), MMEACR precisely captures fine-grained, specific user preferences (highlighted in red) such as a ``cohesive and relaxing mood'' or ``heavy-duty protection with stylish design.'' Furthermore, while AgentCF is vulnerable to interaction noise and suffers from preference drift by Round-5, MMEACR filters cross-modal redundancy via attribute-guided reflection, anchoring user profiles onto stable semantic blocks. This qualitative alignment corroborates our quantitative superiority, visually demonstrating that MMEACR achieves precise long-term profiling through attribute-level memory refinement.


\begin{figure}[t]
    \centering
    \includegraphics[
        width=0.9\columnwidth,
        trim={4pt 4pt 4pt 4pt},
        clip
    ]{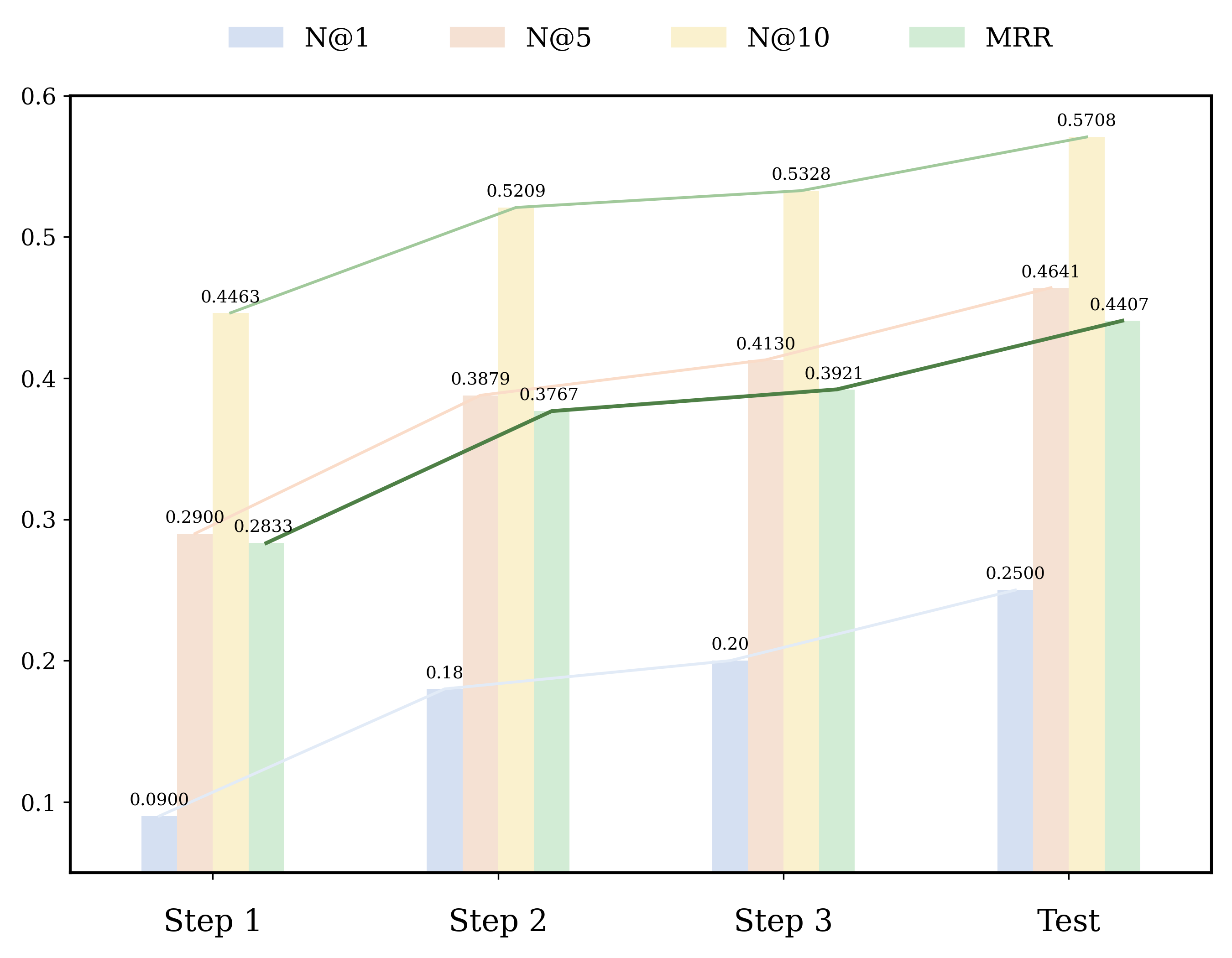}
    \vspace{-0.6em}
    \caption{Performance evolution on Fashion. MMEACR improves steadily across optimization steps and achieves the best results at test time.}
    \label{fig:performance_evolution}
    \vspace{-0.8em}
\end{figure}

\textbf{Ablation Study.}
To evaluate the contribution of each component in our MMEACR framework, we conduct ablation studies by comparing the full \textbf{Multi-AgentCF} model (with both User Agent and Item Agent) against three simplified variants: \textbf{w/o Auto. Interaction}, which removes the automated interaction mechanism between agents and disables iterative mutual refinement; \textbf{w/o User Agent}, which excludes the User Agent and relies solely on item-side modeling; and \textbf{w/o Item Agent}, which removes the Item Agent and retains only user-centric reasoning. Table~\ref{tab:ablation_study} reports results on both cell\_phones and fashion domains. The full Multi-AgentCF consistently achieves the best performance across most metrics, demonstrating the effectiveness of jointly modeling user and item agents within an interactive framework. In contrast, removing either agent leads to clear performance degradation, indicating that both user-side preference modeling and item-side contextual reasoning are essential for accurate ranking. Notably, the largest drop is observed when Auto. Interaction is removed, highlighting the importance of iterative inter-agent communication for refining memory and aligning preferences. These results verify that each component contributes to the overall performance and that their synergistic interaction is crucial for optimal recommendation quality.


\textbf{Effectiveness of Iterative Memory Evolution.}
To investigate how iterative memory refinement enhances recommendation performance, we analyze the evolution of key metrics throughout the optimization process on the Fashion dataset. Specifically, we report results at three consecutive optimization steps and the final testing phase to capture how progressive agent collaboration affects ranking quality over time. As illustrated in Figure~\ref{fig:performance_evolution}, all evaluation metrics, including $N@1$, $N@5$, $N@10$, and $MRR$, exhibit a consistent upward trend as iterative interaction between the user and item agents proceeds. In particular, the $N@10$ score steadily increases from $0.4463$ at Step 1 to $0.5708$ at the final stage, while $MRR$ improves from $0.2833$ to $0.4407$, with similar gains observed for $N@1$ and $N@5$, indicating consistent improvements across different ranking depths. These results suggest that iterative memory refinement enables continuous correction and enrichment of both user preference modeling and item contextual understanding, allowing the system to progressively accumulate more reliable reasoning traces and better align multimodal representations. Overall, the observed performance trajectory validates the effectiveness of our iterative update strategy and highlights the importance of continuous agent collaboration in improving recommendation quality.




\section{Conclusion}
In this paper, we propose MMEACR, a multimodal memory-enhanced agent collaboration framework for recommendation that overcomes modal-lacking limitations in text-only LLM-based systems by jointly leveraging visual and textual features and enabling dynamic memory evolution through iterative agent interactions. By incorporating a dedicated multimodal embedding memory and RRF-based ranking fusion, MMEACR produces more accurate and interpretable recommendations. Extensive experiments on real-world datasets show that MMEACR surpasses strong baselines, particularly in capturing nuanced user preferences from multimodal signals, and ablation studies further validate the contribution of each individual component. In addition, qualitative analyses demonstrate that the proposed agent collaboration mechanism can generate more consistent and preference-aligned reasoning across interaction steps. Overall, our results highlight the effectiveness of integrating multimodal memory with iterative agent collaboration for improving recommendation quality in LLM-based systems.

\section*{Limitations}
While MMEACR demonstrates strong performance and efficient inference, one limitation is the long-term scalability of the continuous memory evolution. As user interactions scale over extended periods, the iterative accumulation in the multimodal embedding memory could potentially increase storage and retrieval costs. Although this is mitigated by our RRF-based ranking fusion in current benchmarks, optimizing the balance between memory capacity and lifelong efficiency remains a promising direction for our future work.

\bibliography{sample-base}

\appendix

\newpage
  
\section{Computational Cost and Efficiency}
Due to the high cost of LLM API calls, we sample 100 users and their historical interactions per dataset, leading to about 500 interaction steps across 5 training rounds. We use a tiered model strategy, where GPT-4o \cite{hurst2024gpt} is applied for memory initialization, interaction simulation, and inference-time ranking, balancing reasoning quality with efficiency and supporting asynchronous batch processing (batch size 4 for training, 5 for evaluation).

At inference, description-based ranking (DES) prompts GPT-4o with the user’s natural-language memory profile and 10 candidate item descriptions to directly output a ranked list, which is mapped back to item IDs using fuzzy matching. In parallel, embedding-based ranking (EMB) encodes user memory and multimodal item content into dense vectors and ranks candidates via cosine similarity without LLM inference. The final result is obtained by fusing DES and EMB rankings using Reciprocal Rank Fusion, with a fallback to EMB-only when DES outputs are invalid.

Overall, inference cost is dominated by a single LLM call per user, and full evaluation remains efficient, completing within about 100 seconds per dataset and showing a 6–16\% speedup over AgentCF due to the streamlined memory design and shorter prompts.

\section{Algorithm}
  \begin{algorithm}[ht]
  \small
  \setstretch{0.9}
  \DontPrintSemicolon
  \SetKwInOut{Input}{Input}
  \SetKwInOut{Output}{Output}
  \SetKwFunction{LLM}{LLM}
  \SetKwFunction{AttrCorr}{Attr$_{+}$}
  \SetKwFunction{AttrInco}{Attr$_{-}$}
  \SetKwFunction{Choice}{ChoicePrompt}
  \SetKwFunction{UpdateUG}{UpdatePrompt}
  \SetKwFunction{Gate}{Gate}
  \SetKwFunction{Match}{Match}

  \caption{Dual-Stage Attribute-Guided Memory Update (UAMG)}

  \Input{Turn $t \in \{1,\dots,T\}$; positive title $s^{+}$, negative title $s^{-}$;
         prior user memory $U^{(t-1)}$; item memories $M^{+}, M^{-}$.}
  \Output{Updated memories $U^{(t)}, M^{+(t)}, M^{-(t)}$.}
  \BlankLine

  \tcp{\textbf{Stage 1: Agent decision}}
  $P_{\text{ch}} \leftarrow \Choice(U^{(t-1)}, s^{+}, s^{-}, M^{+}, M^{-})$\;
  $(\hat{s}, r) \leftarrow \LLM(P_{\text{ch}})$\tcp*{$\hat{s}$: chosen title; $r$: rationale}
  $y \leftarrow \mathbbm{1}[\Match(\hat{s}, s^{+}) > \Match(\hat{s}, s^{-})]$\tcp*{$y=1$: correct}

  \BlankLine
  \tcp{\textbf{Stage 2: Attribute extraction}}
  \eIf{$y = 1$}{
    $A \leftarrow \LLM(\AttrCorr(U^{(t-1)}, s^{+}, s^{-}, M^{+}, M^{-}, r))$\;
  }{
    $A \leftarrow \LLM(\AttrInco(U^{(t-1)}, s^{+}, s^{-}, M^{+}, M^{-}, r))$\;
  }

  \BlankLine
  \tcp{\textbf{Stage 3: Memory update}}
    $(P_u, P_i) \leftarrow \UpdateUG(U^{(t-1)}, M^{+}, M^{-}, s^{+}, s^{-}, r, A, y)$\;
    $\tilde{U}, \tilde{M}^{+}, \tilde{M}^{-} \leftarrow \LLM(P_u),\ \LLM(P_i)$\;

  \Return $U^{(t)}, M^{+(t)}, M^{-(t)}$\;
  \end{algorithm}

\onecolumn
\section{Prompt Templates}\label{app:prompt}

\begin{figure}[H]
\centering
\begin{tcolorbox}[
  colback=green!3,
  colframe=green!30!black,
  arc=6pt,
  boxrule=0.8pt,
  width=\linewidth,
  title=\textbf{Attribute Extraction Templates (Correct Version)},
  coltitle=black,
  colbacktitle=green!25,
  fontupper=\small,
  top=6pt,bottom=6pt,left=8pt,right=8pt
]

\textbf{Use:} Attribute extraction system for user preference analysis. No LLM is used to generate these queries. Each template wraps the deterministic user preferences, item details, and reasoning.

\vspace{0.5em}
\textbf{Variables:}
\begin{itemize}
  \item \texttt{\{user\_description\}}: User preferences description.
  \item \texttt{\{pos\_item\_title\}}: Title of the correct item matching user preference.
  \item \texttt{\{neg\_item\_title\}}: Title of the rejected item.
  \item \texttt{\{system\_reason\}}: Previous system reasoning or explanation.
  \item \texttt{\{attr\_list\}}: Allowed attribute dimensions joined by commas.
\end{itemize}

\vspace{0.5em}
\textbf{Templates:}
\begin{enumerate}
  \item You are an attribute extraction system. Your ONLY task is to extract attributes from the allowed list (\texttt{\{attr\_list\}}) that explain WHY the correct item (\texttt{\{pos\_item\_title\}}) matches the user preference: \texttt{\{user\_description\}}.
  \item Given the correct item \texttt{\{pos\_item\_title\}} and the rejected item \texttt{\{neg\_item\_title\}}, extract at least 2 attributes in the required format based on the user's needs: \texttt{\{user\_description\}}.
  \item Analyze the previous reasoning (\texttt{\{system\_reason\}}) to structure the final answer. Follow STRICT RULES: no explanations, no reasoning, no self-introduction, no markdown, no summaries, and do NOT invent attributes.
\end{enumerate}

\end{tcolorbox}
\caption{The structured query wrapper template containing rules, formatting, and variables for correct attribute analysis and extraction.}
\label{app:prompt:wrappers}
\end{figure}

\begin{figure}[H]
\centering
\begin{tcolorbox}[
  colback=red!3,
  colframe=red!30!black,
  arc=6pt,
  boxrule=0.8pt,
  width=\linewidth,
  title=\textbf{Attribute Extraction Templates (Incorrect Version)},
  coltitle=black,
  colbacktitle=red!25,
  fontupper=\small,
  top=6pt,bottom=6pt,left=8pt,right=8pt
]

\textbf{Use:} LLM-based prompt templates for simultaneous positive/negative attribute extraction. This setup forces the model to perform complex evaluation, scoring, and extreme negative filtering within a single completion step.

\vspace{0.5em}
\textbf{Variables:}
\begin{itemize}
  \item \texttt{\{user\_description\}}: User preferences description.
  \item \texttt{\{pos\_item\_title\}}: Title of the correct item matching user preference.
  \item \texttt{\{neg\_item\_title\}}: Title of the rejected item.
  \item \texttt{\{system\_reason\}}: Previous system reasoning or explanation.
  \item \texttt{\{attr\_list\}}: Allowed attribute dimensions joined by commas.
\end{itemize}

\vspace{0.5em}
\textbf{Templates:}
\begin{enumerate}
  \item You are an attribute extraction system. Your ONLY task is to extract positive attributes from \texttt{\{pos\_item\_title\}} and misleading negative attributes from \texttt{\{neg\_item\_title\}} that explain the user preference: \texttt{\{user\_description\}}. Allowed attributes ONLY: \texttt{\{attr\_list\}}.
  
  \item Given the correct item \texttt{\{pos\_item\_title\}} and the rejected item \texttt{\{neg\_item\_title\}}, extract at least 2 attributes in the required format based on the user's needs: \texttt{\{user\_description\}}.
  
  \item Analyze the previous reasoning (\texttt{\{system\_reason\}}) to structure the final answer. Follow STRICT RULES: no explanations, no reasoning, no self-introduction, no markdown, no summaries, and do NOT invent attributes.
\end{enumerate}

\end{tcolorbox}
\caption{The structured query wrapper template containing rules, formatting, and variables for incorrect attribute analysis and extraction.}
\label{app:prompt:wrappers_incorrect}
\end{figure}

\begin{figure}[H]
\centering
\begin{tcolorbox}[
  colback=green!3,
  colframe=green!30!black,
  arc=6pt,
  boxrule=0.8pt,
  width=\linewidth,
  title=\textbf{User Memory Reflection with attribute guidance Template (Correct Version)},
  coltitle=black,
  colbacktitle=green!25,
  fontupper=\small,
  top=6pt,bottom=6pt,left=8pt,right=8pt
]

\textbf{Use:} LLM-based prompt template for validating correct preference judgments through attribute-level rationale extraction and personalized self-introduction refinement. The template encourages the model to reinforce accurate preferences and consolidate stable user characteristics.

\vspace{0.5em}
\textbf{Variables:}
\begin{itemize}
  \item \texttt{\{list\_of\_item\_description\}}: Descriptions of the compared candidate items.
  \item \texttt{\{pos\_item\_title\}}: Title of the correctly selected item preferred by the user.
  \item \texttt{\{neg\_item\_title\}}: Title of the rejected item disliked by the user.
  \item \texttt{\{system\_reason\}}: User’s previous reasoning or explanation for the successful choice.
  \item \texttt{\{attribute\_dimensions\}}: Allowed attribute dimensions used for attribute-level preference analysis.
\end{itemize}

\vspace{0.5em}
\textbf{Templates:}
\begin{enumerate}
  \item Recently, you compared two candidate items described in \texttt{\{list\_of\_item\_description\}}. You selected \texttt{\{pos\_item\_title\}} and rejected \texttt{\{neg\_item\_title\}}. Your explanation was: \texttt{\{system\_reason\}}.
  
  \item After encountering these items, you confirmed that you truly prefer \texttt{\{pos\_item\_title\}} and dislike \texttt{\{neg\_item\_title\}}. This indicates your original judgment about your preferences was accurate.
  
  \item Perform attribute-level rationale extraction using ONLY the allowed dimensions: \texttt{\{attribute\_dimensions\}}. Identify 1--3 key attributes that contributed to this successful match. For each attribute, specify:
  \begin{itemize}
      \item the related item,
      \item whether the attribute is positive or negative to the preference,
      \item and an importance score from 1 to 5.
  \end{itemize}
  
  \item Based on the extracted attributes, analyze and reinforce:
  \begin{itemize}
      \item confirmed preferences from \texttt{\{pos\_item\_title\}},
      \item confirmed dislikes from \texttt{\{neg\_item\_title\}},
      \item and stable preference patterns reflected in the previous reasoning.
  \end{itemize}
  
  \item Generate an updated self-introduction by:
  \begin{itemize}
      \item first emphasizing newly confirmed preferences,
      \item then summarizing consistent past preferences,
      \item and finally describing dislikes.
  \end{itemize}
  
  \item Follow STRICT RULES:
  \begin{itemize}
      \item Output MUST strictly follow the required structure.
      \item The self-introduction must be under 150 words.
      \item Be concise, specific, and personalized.
      \item Do NOT include reasoning processes in the self-introduction.
      \item Avoid generic preference descriptions.
  \end{itemize}
\end{enumerate}

\end{tcolorbox}
\caption{The structured prompt template for successful preference confirmation and personalized self-introduction refinement.}
\label{app:prompt:successful_preference_template}
\end{figure}

\begin{figure}[H]
\centering
\begin{tcolorbox}[
  colback=red!3,
  colframe=red!30!black,
  arc=6pt,
  boxrule=0.8pt,
  width=\linewidth,
  title=\textbf{User Memory Reflection with attribute guidance Template (Incorrect Version)},
  coltitle=black,
  colbacktitle=red!25,
  fontupper=\small,
  top=6pt,bottom=6pt,left=8pt,right=8pt
]

\textbf{Use:} LLM-based prompt template for correcting mistaken preference judgments through attribute-level reflection and self-introduction updating. The template guides the model to analyze misconception attributes, identify true preferences/dislikes, and refine personalized user profiles.

\vspace{0.5em}
\textbf{Variables:}
\begin{itemize}
  \item \texttt{\{list\_of\_item\_description\}}: Descriptions of the compared candidate items.
  \item \texttt{\{pos\_item\_title\}}: Title of the finally preferred (correct) item.
  \item \texttt{\{neg\_item\_title\}}: Title of the mistakenly chosen and later rejected item.
  \item \texttt{\{system\_reason\}}: User’s previous reasoning or explanation for the incorrect choice.
  \item \texttt{\{attribute\_dimensions\}}: Allowed attribute dimensions used for attribute-level analysis.
\end{itemize}

\vspace{0.5em}
\textbf{Templates:}
\begin{enumerate}
  \item Recently, you compared two candidate items described in \texttt{\{list\_of\_item\_description\}}. Although you initially selected \texttt{\{neg\_item\_title\}}, you later realized that \texttt{\{pos\_item\_title\}} better matches your true preferences. Your previous reasoning was: \texttt{\{system\_reason\}}.
  
  \item Perform attribute-level misconception analysis using ONLY the allowed dimensions: \texttt{\{attribute\_dimensions\}}. Identify 1--3 key attributes that caused the mistaken judgment. For each attribute, specify:
  \begin{itemize}
      \item the related item,
      \item whether the attribute is positive or negative to the true preference,
      \item and an importance score from 1 to 5.
  \end{itemize}
  
  \item Based on the extracted attributes, analyze why the previous judgment was incorrect and identify:
  \begin{itemize}
      \item newly discovered preferences from \texttt{\{pos\_item\_title\}},
      \item newly identified dislikes from \texttt{\{neg\_item\_title\}},
      \item and conflicts between old and updated preferences.
  \end{itemize}
  
  \item Generate an updated self-introduction by:
  \begin{itemize}
      \item first emphasizing new preferences,
      \item then summarizing consistent past preferences,
      \item and finally describing dislikes.
  \end{itemize}
  
  \item Follow STRICT RULES:
  \begin{itemize}
      \item Output MUST strictly follow the required structure.
      \item The self-introduction must be under 150 words.
      \item Be concise, specific, and personalized.
      \item Do NOT include reasoning processes in the self-introduction.
      \item Avoid generic preference descriptions.
  \end{itemize}
\end{enumerate}

\end{tcolorbox}
\caption{The structured prompt template for attribute-level misconception correction and personalized self-introduction updating.}
\label{app:prompt:preference_correction_template}
\end{figure}

\begin{figure}[H]
\centering
\begin{tcolorbox}[
  colback=green!3,
  colframe=green!30!black,
  arc=6pt,
  boxrule=0.8pt,
  width=\linewidth,
  title=\textbf{User Memory Reflection without attribute guidance Template (Correct Version)},
  coltitle=black,
  colbacktitle=green!25,
  fontupper=\small,
  top=6pt,bottom=6pt,left=8pt,right=8pt
]

\textbf{Use:} LLM-based prompt template for reinforcing correct preference judgments without explicit attribute-level guidance. The template helps the model consolidate accurate user preferences and refine personalized self-introductions based on successful interaction outcomes.

\vspace{0.5em}
\textbf{Variables:}
\begin{itemize}
  \item \texttt{\{list\_of\_item\_description\}}: Descriptions of the compared candidate items.
  \item \texttt{\{pos\_item\_title\}}: Title of the correctly selected and preferred item.
  \item \texttt{\{neg\_item\_title\}}: Title of the rejected and disliked item.
  \item \texttt{\{system\_reason\}}: User’s previous reasoning or explanation for the successful choice.
\end{itemize}

\vspace{0.5em}
\textbf{Templates:}
\begin{enumerate}
  \item Recently, you compared two candidate items described in \texttt{\{list\_of\_item\_description\}}. You selected \texttt{\{pos\_item\_title\}} and rejected \texttt{\{neg\_item\_title\}}. Your previous reasoning was: \texttt{\{system\_reason\}}.
  
  \item After encountering these items, you confirmed that you truly prefer \texttt{\{pos\_item\_title\}} and dislike \texttt{\{neg\_item\_title\}}. This indicates your original judgment about your preferences was accurate.
  
  \item Analyze your previous reasoning and identify:
  \begin{itemize}
      \item preference patterns correctly reflected in your earlier judgment,
      \item newly confirmed preferences from \texttt{\{pos\_item\_title\}},
      \item and dislikes associated with \texttt{\{neg\_item\_title\}}.
  \end{itemize}
  
  \item Summarize your previous preferences and dislikes, combining them with newly confirmed insights while removing inconsistent or conflicting parts.
  
  \item Generate an updated self-introduction by:
  \begin{itemize}
      \item first emphasizing newly confirmed preferences,
      \item then summarizing consistent past preferences,
      \item and finally describing dislikes.
  \end{itemize}
  
  \item Follow STRICT RULES:
  \begin{itemize}
      \item Output format MUST be:
      \texttt{My updated self-introduction: [Your updated self-introduction here].}
      \item The self-introduction must be under 150 words.
      \item Be concise, clear, and personalized.
      \item Do NOT describe reasoning processes in the self-introduction.
      \item Describe only preferred or disliked item characteristics.
      \item Avoid generic preference descriptions.
  \end{itemize}
\end{enumerate}

\end{tcolorbox}
\caption{The structured prompt template for successful preference confirmation and self-introduction refinement without explicit attribute-level guidance.}
\label{app:prompt:successful_preference_no_attribute}
\end{figure}

\begin{figure}[H]
\centering
\begin{tcolorbox}[
  colback=red!3,
  colframe=red!30!black,
  arc=6pt,
  boxrule=0.8pt,
  width=\linewidth,
  title=\textbf{User Memory Reflection without attribute guidance Template (Incorrect Version)},
  coltitle=black,
  colbacktitle=red!25,
  fontupper=\small,
  top=6pt,bottom=6pt,left=8pt,right=8pt
]

\textbf{Use:} LLM-based prompt template for correcting mistaken preference judgments without explicit attribute-level guidance. The template directly guides the model to revise preference understanding and update personalized self-introductions based on corrected user experiences.

\vspace{0.5em}
\textbf{Variables:}
\begin{itemize}
  \item \texttt{\{list\_of\_item\_description\}}: Descriptions of the compared candidate items.
  \item \texttt{\{pos\_item\_title\}}: Title of the finally preferred (correct) item.
  \item \texttt{\{neg\_item\_title\}}: Title of the mistakenly selected and later disliked item.
  \item \texttt{\{system\_reason\}}: User’s previous reasoning or explanation for the incorrect choice.
\end{itemize}

\vspace{0.5em}
\textbf{Templates:}
\begin{enumerate}
  \item Recently, you compared two candidate items described in \texttt{\{list\_of\_item\_description\}}. Although you initially selected \texttt{\{neg\_item\_title\}}, you later realized that \texttt{\{pos\_item\_title\}} better matches your true preferences. Your previous reasoning was: \texttt{\{system\_reason\}}.
  
  \item Analyze the misconceptions in your previous judgment and explain how your understanding of your preferences has changed after encountering these items.
  
  \item Identify:
  \begin{itemize}
      \item new preferences discovered from \texttt{\{pos\_item\_title\}},
      \item new dislikes identified from \texttt{\{neg\_item\_title\}},
      \item and conflicts between old and updated preferences.
  \end{itemize}
  
  \item Summarize your previous preferences and merge them with the newly discovered insights while removing inconsistent or conflicting parts.
  
  \item Generate an updated self-introduction by:
  \begin{itemize}
      \item first emphasizing new preferences,
      \item then summarizing past consistent preferences,
      \item and finally describing dislikes.
  \end{itemize}
  
  \item Follow STRICT RULES:
  \begin{itemize}
      \item Output format MUST be:
      \texttt{My updated self-introduction: [Your updated self-introduction here].}
      \item The self-introduction must be under 150 words.
      \item Be concise, clear, and personalized.
      \item Do NOT describe reasoning processes in the self-introduction.
      \item Describe only preferred or disliked item characteristics.
      \item Avoid generic preference descriptions.
  \end{itemize}
\end{enumerate}

\end{tcolorbox}
\caption{The structured prompt template for preference correction and self-introduction updating without explicit attribute-level guidance.}
\label{app:prompt:preference_correction_no_attribute}
\end{figure}

\end{document}